\documentclass[prl,aps,twocolumn,preprintnumbers,amsmath,amssymb,superscriptaddress,showpacs]{revtex4}

\usepackage{graphicx}
\usepackage{dcolumn}
\usepackage{bm}
\usepackage{epstopdf}

\newcount\driver
\newcount\bozza

scaled\magstep1                  
scaled\magstep1 
 
scaled\magstep1                 \font\indbf=cmbx10 scaled\magstep2

scaled \magstep2

{\count255=\time\divide\count255 by 60
\xdef\hourmin{\number\count255}
        \multiply\count255 by-60\advance\count255 by\time
   \xdef\hourmin{\hourmin:\ifnum\count255<10 0\fi\the\count255}}

\let\a=\alpha \let\b=\beta    \let\g=\gamma     \let\d=\delta     \let\e=\varepsilon
  \let\h=\eta           \let\l=\lambda
\let\m=\mu    \let\n=\nu              \let\p=\pi        
\let\s=\sigma             
\let\ps=\psi   \let\o=\omega     
        \let\L=\Lambda    
             
\let\O=\Omega

\def\VV{{\cal V}}

\def\DD{{\cal D}}

\def\pp{{\bf p}}\def\qq{{\bf q}}\def\xx{{\bf x}}
   
\def\yy{{\bf y}}\def\kk{{\bf k}}\def\nn{{\bf n}}

       \def\oo{{\underline \omega}}
\def\ee{{\underline \varepsilon}}

          \def\BBB{\hbox{\euftw B}}

\let\==\equiv

\let\io=\infty
\let\0=\noindent

\def\*{{\hfill\break\null\hfill\break}}

\def\tilde#1{{\widetilde #1}}

\def\tende#1{\,\vtop{\ialign{##\crcr\rightarrowfill\crcr
             \noalign{\kern-1pt\nointerlineskip}
             \hskip3.pt${\scriptstyle #1}$\hskip3.pt\crcr}}\,}
\def\otto{\,{\kern-1.truept\leftarrow\kern-5.truept\to\kern-1.truept}\,}

\def\wh#1{\widehat{#1}}
\def\hat#1{\wh{#1}}
\def\sqt[#1]#2{\root #1\of {#2}}

\def\bp{{\bar \ps}}

\def\VV{{\cal V}}

\def\DD{{\cal D}}

\def\T#1{{#1_{\kern-3pt\lower7pt\hbox{$\widetilde{}$}}\kern3pt}}
\def\VVV#1{{\underline #1}_{\kern-3pt
\lower7pt\hbox{$\widetilde{}$}}\kern3pt\,}
\def\W#1{#1_{\kern-3pt\lower7.5pt\hbox{$\widetilde{}$}}\kern2pt\,}

\def\indica{\leaders \hbox to 0.5cm{\hss.\hss}\hfill}
\def\guida{\leaders\hbox to 1em{\hss.\hss}\hfill}
\mathchardef\oo= "0521

\def\pp{{\bf p}}\def\qq{{\bf q}}\def\xx{{\bf x}}
\def\yy{{\bf y}}\def\kk{{\bf k}}\def\nn{{\bf n}}

\def\oo{{\underline \omega}}

\def\qed{\raise1pt\hbox{\vrule height5pt width5pt depth0pt}}
 
  \def\bp{{\bar p}} 

\def\indic{\hbox{\raise-2pt \hbox{\indbf 1}}}

\def\ins#1#2#3{\vbox to0pt{\kern-#2 \hbox{\kern#1 #3}\vss}\nointerlineskip}

\newdimen\xshift \newdimen\xwidth \newdimen\yshift
\newcount\griglia

\def\insertplot#1#2#3#4#5#6{%
\xwidth=#1pt \xshift=\hsize \advance\xshift by-\xwidth \divide\xshift by 2%
\begin{figure}[ht]
\vspace{#2pt} \hspace{\xshift}
\begin{minipage}{#1pt}
#3 \ifnum\driver=1 \griglia=#6
\ifnum\griglia=1 \openout13=griglia.ps \write13{gsave .2
setlinewidth} \write13{0 10 #1 {dup 0 moveto #2 lineto } for}
\write13{0 10 #2 {dup 0 exch moveto #1 exch lineto } for}
\write13{stroke} \write13{.5 setlinewidth} \write13{0 50 #1 {dup 0
moveto #2 lineto } for} \write13{0 50 #2 {dup 0 exch moveto #1
exch lineto } for} \write13{stroke grestore} \closeout13
\includegraphics{griglia.ps} \fi
\includegraphics{#4.ps}\fi%
\ifnum\driver=2 \fi
\end{minipage}
\caption{#5}
\end{figure}
}

\newdimen\shift \shift=-1.5truecm
\def\lb#1{%
\ifnum\bozza=1
\label{#1}\rlap{\hbox{\hskip\shift$\scriptstyle#1$}}
\else\label{#1} \fi}

\def\be{\begin{equation}}
\def\ee{\end{equation}}
\def\bea{\begin{eqnarray}}\def\eea{\end{eqnarray}}
\def\bean{\begin{eqnarray*}}\def\eean{\end{eqnarray*}}
\def\bfr{\begin{flushright}}\def\efr{\end{flushright}}
\def\bc{\begin{center}}\def\ec{\end{center}}
\def\bal{\begin{align}}\def\eal{\end{align}}
\def\ba#1{\begin{array}{#1}} \def\ea{\end{array}}
\def\bd{\begin{description}}\def\ed{\end{description}}
\def\bv{\begin{verbatim}}\def\ev{\end{verbatim}}
\def\nn{\nonumber}
\def\Halmos{\hfill\vrule height10pt width4pt depth2pt \par\hbox to \hsize{}}
\def\pref#1{(\ref{#1})}

\def\ins#1#2#3{\vbox to0pt{\kern-#2 \hbox{\kern#1 #3}\vss}\nointerlineskip}

\newdimen\xshift \newdimen\xwidth \newdimen\yshift
\newcount\griglia

\def\insertplot#1#2#3#4#5#6{%
\xwidth=#1pt \xshift=\hsize \advance\xshift by-\xwidth \divide\xshift by 2%
\begin{figure}[ht]
\vspace{#2pt} \hspace{\xshift}
\begin{minipage}{#1pt}
#3 \ifnum\driver=1 \griglia=#6
\ifnum\griglia=1 \openout13=griglia.ps \write13{gsave .2
setlinewidth} \write13{0 10 #1 {dup 0 moveto #2 lineto } for}
\write13{0 10 #2 {dup 0 exch moveto #1 exch lineto } for}
\write13{stroke} \write13{.5 setlinewidth} \write13{0 50 #1 {dup 0
moveto #2 lineto } for} \write13{0 50 #2 {dup 0 exch moveto #1
exch lineto } for} \write13{stroke grestore} \closeout13
\includegraphics{griglia.ps} \fi
\includegraphics{#4.ps}\fi%
\ifnum\driver=2 \fi
\end{minipage}
\caption{#5}
\end{figure}
}

\newdimen\shift \shift=-1.5truecm
\def\lb#1{%
\label{#1}\rlap{\hbox{\hskip\shift$\scriptstyle#1$}}
\else\label{#1} \fi}

\def\be{\begin{equation}}
\def\ee{\end{equation}}
\def\bea{\begin{eqnarray}}\def\eea{\end{eqnarray}}
\def\bean{\begin{eqnarray*}}\def\eean{\end{eqnarray*}}
\def\bfr{\begin{flushright}}\def\efr{\end{flushright}}
\def\bc{\begin{center}}\def\ec{\end{center}}
\def\bal{\begin{align}}\def\eal{\end{align}}
\def\ba#1{\begin{array}{#1}} \def\ea{\end{array}}
\def\bd{\begin{description}}\def\ed{\end{description}}
\def\bv{\begin{verbatim}}\def\ev{\end{verbatim}}
\def\nn{\nonumber}
\def\Halmos{\hfill\vrule height10pt width4pt depth2pt \par\hbox to \hsize{}}
\def\pref#1{(\ref{#1})}

\usepackage{amsmath}
\usepackage{amsfonts}
\usepackage{amssymb}
\usepackage{slashbox}
\usepackage{epstopdf}

scaled\magstep1

\let\a=\alpha \let\b=\beta  \let\g=\gamma  \let\d=\delta
\let\e=\varepsilon
  \let\h=\eta     \let\l=\lambda
\let\m=\mu    \let\n=\nu         \let\p=\pi    
\let\s=\sigma     
\let\ps=\Psi   \let\o=\omega
   \let\L=\Lambda 
         
\let\O=\Omega 

 \def\VV{{\cal V}}
 
 \def\BBB{{\cal B}}
  
\def\DD{{\cal D}}

\def\qq{{\bf q}} \def\pp{{\bf p}}
 \def\xx{{\bf x}} \def\yy{{\bf y}} 
\def\kk{{\bf k}}

\def\nn{\nonumber}

\def\\{\hfill\break}
\def\={:=}
\let\io=\infty
\let\0=\noindent

\def\tende#1{\,\vtop{\ialign{##\crcr\rightarrowfill\crcr\noalign{\kern-1pt
    \nointerlineskip} \hskip3.pt${\scriptstyle #1}$\hskip3.pt\crcr}}\,}
\def\otto{\,{\kern-1.truept\leftarrow\kern-5.truept\to\kern-1.truept}\,}

\def\wh{\widehat}
\def\to{\rightarrow}

\def\qed{\hfill\raise1pt\hbox{\vrule height5pt width5pt depth0pt}}

\def\be{\begin{equation}}
\def\ee{\end{equation}}
\def\bp{\begin{pmatrix}}
\def\ep{\end{pmatrix}}
\def\bea{\begin{eqnarray}}
\def\eea{\end{eqnarray}}
\def\nn{\nonumber}
\def\pref#1{(\ref{#1})}

\def\lb{\label}

\begin{document}

\title{Universal conductivity of graphene in the ultra-relativistic regime}

\author{Igor F. Herbut}
\affiliation{ Department of Physics, Simon Fraser University,
Burnaby, British Columbia, Canada V5A 1S6} \affiliation{
Max-Planck-Institut f\"ur Physik Komplexer Systeme, N\"othnitzer
Strasse 38, 01187 Dresden, Germany}

\author{Vieri Mastropietro}
\affiliation{Departimento di Matematica ``Federigo Enriques",
Universit\`a degli Studi di Milano, Via Cesare Saldini 50, 20133
Milano, Italy}

\

\begin{abstract}
We calculate the optical ($\Lambda\gg \omega \gg T$)  conductivity in clean graphene in the ultimate low-energy regime, when retardation effects of the electromagnetic interaction become important
and when the full Lorentz symmetry emerges. In contrast to what happens with the short range or with the Coulomb long-range instantaneous interactions,
the optical conductivity is now no longer equal to its non interacting value, but acquires universal corrections in powers of the fine structure constant. The coefficient of the first order correction is computed, and found to be of order one. We also present the result for the conductivity in the large-N limit, with $N$ as the number of Dirac fermions species, to the order $1/N^2$.
\end{abstract}

\maketitle

\section{Introduction}

Graphene owes several of its remarkable properties to the fact
that it admits an effective relativistic quantum field theory
description in terms of massless Dirac fermions in two dimension.
A dramatic manifestation of this fact is seen  in its conductivity
properties; recent experiments \cite{N1} found
that the optical conductivity in monolayer graphene is essentially
constant in a wide range of frequencies, and very close to the
value $(\pi/2)(e^2/h)$ (that is $1/4$ in the natural units
$e=\hbar=c=1$ which we will use), which also happens to be the value
found for the system of non-interacting Dirac fermions at half filling \cite{L}.
This remarkable result, however, raises a couple of natural questions of principle: why
the interactions, which, at least when taken at face value, are not particularly weak in graphene, do not
produce visible corrections to the non interacting value of the
conductivity?  And should the conductivity, and in the optical limit in particular, in graphene be {\it in principle} 
equal to its non interacting value, or there are many body corrections which may lie inside the experimental errors?

The computation of the many body interaction effects on graphene's
conductivity is quite sensitive to regularizations and
approximations, and several aspects of it have been controversial.
In the case of {\it short range} interactions, after first
perturbative computations claiming non vanishing corrections, it
was finally rigorously proved \cite{GMPcond} that there are {\it
no interaction corrections} to the  (zero temperature) zero frequency conductivity; all the
interaction contributions to the conductivity cancel out at all
orders in the renormalized expansion. The exact vanishing of
interaction correction emerges as a consequence of the Ward
identities and the irrelevance, in the technical renormalization
group (RG) sense, of the interaction,\cite{GMPcond, H1}. On the
other hand, in the case of {\it long range} Coulomb interactions
it has been predicted that the optical conductivity is still universal
and equal to $1/4$ \cite{H0}, the argument this time being based
on the divergence of the Fermi velocity \cite{V3}, and the {\it
relative} (to the kinetic energy) irrelevance of the interaction.
The low frequency correction to the conductivity was found to be
not particularly small, due to the slow logarithmic increase of
the Fermi velocity, which is what is expected in absence of
(possible) accidental cancellation. A controversy in the
computation of such corrections arose  in literature,
\cite{H0,SS2,M2,H1,SF}. Technically the reason for the controversy
lies in the ambiguities produced by the ultraviolet divergences
due to the continuum limit. Recently, however, the controversy has
been claimed to be settled by performing a lattice computation in
\cite{RL},  in favor of the value originally found in \cite{H1}.

On the other hand, the Fermi velocity divergence found in the Coulomb case at very low frequencies is clearly rather unphysical,
and simply signals ultimate inadequacy of the usual model of instantaneous  Coulomb interaction. With the increase of the Fermi velocity the
retardation effects eventually become important, so that the retarded current-current interaction must be added to the
Coulomb density-density interaction. Such effects have been analyzed before in
\cite{GGV} , \cite{GMPgauge} by a RG analysis, and it was found that
the flow of the Fermi velocity stops at the velocity of light $c$, and, maybe most importantly,
that the coupling constant (i. e. the charge) in the theory is {\it exactly} marginal.
In particular, a lattice model for graphene interacting with an electromagnetic field was considered in \cite{GMPgauge},
and by iterating the RG it was proved that the Lorentz symmetry spontaneously emerges, and that the system is asymptotically close
to the so-called ``reduced" quantum electrodynamics ($QED_{4,3}$).

The aim of this paper is to compute the corrections to the non-interacting value of the universal optical conductivity in the limit $\omega\gg T$, 
due to the {\it full} electromagnetic interaction. We show first by exact RG methods that, with the retardation effects
and with the honeycomb lattice included, in units of $e^2/\hbar$,
\begin{equation}\label{aa}
\sigma= \frac{N}{8} ( 1+ (C_1 - \frac{N}{8}) {e^2\over 2 } + O(e^4)),
\end{equation}
where $N$ is the number of four-component Dirac fermions ($N=2$ in graphene), and $C_1$ is an $N$-independent constant.
In the ultimate, low-energy regime the conductivity is therefore {\it different} with respect to its non interacting vale of $1/4$;
the dependence on the number of fermion components $N$ in the first interaction correction immediately rules out possible cancellations in general.
On the other hand, the correction to conductivity, in vacuum with the dielectric constant $\epsilon=1$ assumed above, is still {\it universal},
as a consequence of the emerging Lorentz invariance, in the sense that it does not depend on the material parameters such as the Fermi velocity, but only on the
fine structure constant.

As the lattice model for graphene becomes asymptotically close to $QED_{4,3}$, which is the fixed point of the RG flow,
it is worthwhile computing the optical conductivity
directly in this continuum model by using standard field-theoretical methods.
This way an expression identical to \pref{aa} is found, with
the numerical constant $C_1= 0.0089319$. The leading $O(e^2 N)$ correction is {\it identical} both in the lattice and in the continuum model. In graphene, the number of Dirac fermions is $N=2$, and this correction is $0.125$ and much larger than the constant $C_1$. In any case,
as $e^2 = 4\pi \alpha $, with $\alpha=1/137.036$ as the fine structure constant, in vacuum we find
$\s/ \s_0=1-0.01064$, a small correction, and at the moment within  the experimental error in
 \cite{N1}. The conductivity  is therefore {\it in principle}, if not in practice, different from the non interacting value and, because of the smallness of the fine structure constant, with a new universal value only slightly reduced from the non-interacting one.
We may also observe in passing that the prefactor multiplying the charge $e^2 /2$ in the correction term is $0.11607$, of the same order in magnitude as found in the non-relativistic limit for the static Coulomb interaction \cite{H0,H1,RL}, although of the opposite sign.

Finally, possible dynamical effects of the electron spin, such as the opening of the small gap at the Dirac point due to the spin-orbit interaction,\cite{kane} will in this paper, just as in all the previous work,\cite{GGV,GMPgauge} be neglected.

\section{Renormalization Group computation}

We can describe graphene by a system of
electrons on the honeycomb lattice interacting with an
electromagnetic  field. The Hamiltonian is
$H=H_e+H_a$, where
\be H_e=-t\sum_{\vec x\in\L\atop j=1,2,3}\sum_{\s=1}^N a^+_{\vec
x,\s}b^-_{\vec x+\vec \d_i,\s} e^{ie\int_0^1 \vec \d_j \vec a(\vec
x+s\vec \d_j,0)}+c.c.\label{asa}, \ee
with $\L=(n_1 \vec l_1+\vec n_2 \vec l_2)$,
$n_i=0,..,L-1$ and $\vec l_{1,2}= \frac12(3,\pm\sqrt3)$,
$\vec\d_1=(1,0)$, $\vec\d_2={1\over 2}(-1,\sqrt{3})$, $\vec\d_3={1\over 2}(-1,-\sqrt{3})$
and $H_a$ is the free photon Hamiltonian.

The physical observables are conveniently obtained in terms of the following generating functional for the external source fields $A$ and $\lambda$:
\be
e^{W_{L,\b}(A,\l)}=\int P(d\psi)\int
P(da)e^{\VV(a+A,\psi)+(\psi,\l)}\label{por},
\ee
where $\psi^\pm_{\xx}=(a^\pm_{\xx,\s},b^\pm_{\xx+\d_1,\s})$ are
Grassman variables (denoted with a slight abuse of notation with
the same symbol), $\s=1,..,N$, $\d_{j}=(0,\vec\d_j)$, $P(d\psi)$
is the fermionic integration with  propagator
\be
g(\xx-\yy)={1\over \b |\L|}\sum_{\kk\in \DD}e^{i\kk(\xx-\yy)}
\left(\begin{array}{cc}
ik_0 & v \O^*(\vec k)
\\ v \O(\vec k) & ik_0 \end{array}\right)^{-1}\;\label{vo}\ee
with $\xx=(x_0,\vec x)$, $\kk=(k_0,\vec k)$, $k_{0}={2\pi\over \b}(m+1/2)$,
$\vec k={m_1\over L}\vec
b_1+{m_2\over L}\vec b_2$,  $\vec b_{1,2}={2\pi\over 3}(1,\pm \sqrt{3})$, $0\le m_i<L$,
$|\L|=L^2$,
$v={3\over 2}t$,
and $\O(\vec k) = \frac23 \sum_{j=1,2,3}e^{i\vec k(\vec \d_j -
\vec\d_1)}$.
In the limit $L\to\io$ ${1\over |\L|}\sum_{\vec k}\to S
\int_{\BBB}\frac{d\vec k}{(2\pi)^2}$, $S={3\sqrt{3}\over 2}$ is the area of the exhagonal cell
and the integral is over
the Brillouin zone.
The dispersion relation  $\O(\vec k)$ vanishes at the two Fermi points $\kk^\pm_F=(0,{2\pi\over 3},\pm {2\pi\over 3\sqrt{3}})$.
$a_{\m,\xx}$ is a Gaussian variable while $P(da)$ is the photon integration with propagator
$\d_{\m,\n} w(\xx-\yy)$ and
$\hat w(\pp)={1\over S}  {\chi(\pp)\over 2|\pp|}$,
where $\chi(\pp)$ is a cut-off functions and the Feynman gauge is assumed. Finally the interaction $\VV$
can be easily deduced from \pref{asa} and it has the form
\be
V(a,\psi)=\sum_{\m=1,2,3}e \int d\xx (a_{0,\xx} j_{0,\xx}+ v \vec a_{\xx} \vec j_{\xx})+ F(a),
\ee
where $F(a)$ contains higher order term in $a$, $j_{\m,\xx}=(j_{0,\xx},\vec j_{\xx})_\m$ with
\bea
&&j_{0,\xx}=\sum_{\s=1}^N[ a^+_{\xx,\s}a^-_{\xx,\s}+a^+_{\xx+{\bf \d}_1,\s}a^-_{\xx+{\bf\d}_1,\s}]\nn\\
&&\vec j_{\xx}={2\over 3}\sum_{j=1,2,3}\sum_{\s=1}^N \vec \d_j (a^+_{\xx,\s}b^-_{\xx+\d_j,\s}-b^+_{\xx+\d_j,\s}
a^-_{\xx,\s}).
\eea
From the identity $
W_{L,\b}(A,\l)=W_{L,\b}(A+\partial\a,\l e^{i\a_\xx})$ we get
\be
{\partial\over\partial \a} W_{L,\b}(A+\partial\a,\l e^{i\a_\xx})=0\label{wi},
\ee
and the derivatives of the above relations provide a set of Ward identities. Calling
$K_{\m,\n}(\pp)={1\over S}
{\partial^2 W_{L,\b}(A,0)\over\partial\hat A_{\m,\pp}\partial\hat A_{\n,-\pp}}|_{0}$, $\pp=(\o,p)$,
the zero-temperature conductivity, at zero frequency, is defined via Kubo formula
$\s=\lim_{\o\to 0} -e^2{1\over \o}
K_{i,i}(\o,0)$.
Note that $K_{i,j}(\pp)$ is the sum of the truncated current-current correlation and of the diamagnetic term,
which is a constant in $\pp$; by the Ward identity obtained from \pref{wi} with a derivative with respect to $A$,
we get
$
\lim_{p_1\to 0}\lim_{\o\to 0}\hat K_{11}(\pp)|_{p_2=0}=\lim_{p_1\to 0}\lim_{\o\to 0}{\o\over p_1} \hat K_{01}(\pp)=0
$
and,
as $\hat K_{11}(\pp)$ is continuous at weak coupling, we can reverse the limits so that the conductivity can be written as
\be \s=\lim_{\o\to 0^+ } -e^2\frac{\hat K_{ii}(\o,0)-\hat K_{ii}(0,0)}{\o} \; \label{conc}.
\ee
There is therefore no need of computing the diamagnetic term, but it is sufficient to compute the current-current correlation and subtract the value in $\pp=(0,0)$.

$K_{i,j}(\pp)$ can be computed by Wilsonian RG; for details,  see
\cite{GMPgauge}. The starting point consists of writing the Grassman variables
as sums of variables with momenta closer and closer to the two Fermi points
$\kk^\pm_F=(0,{2\pi\over 3},\pm {2\pi\over 3\sqrt{3}})$, that is
\be \hat\psi=\psi^{(1)}+\sum_{\e=\pm} \sum_{h=-\io}^0 \psi^{(h)}_{\e}\ee where $\psi^{(h)}_\e$
lives on a shell of momenta distant $O(2^h)$ from $\kk^\e_F$.
Similarly,  we can write $a^\m= \sum_{h=-\io}^0 a^\m_h$, and
$a^\m_h$ has propagator $f_h(\kk) w(\kk)$ with $f_h(\kk)$ non vanishing for $2^{h-1}\le |\kk|\le 2^{h+1}$.
After the integration of the fields $(\psi^{1},a^1),..(\psi^{h},a^h)$,  one finds that the generating functional can be written as
\begin{widetext}
\bea e^{W_{L,\b}(A,\l)}= e^{B^h(A,\l)}\int \prod_{\e=\pm 1}
P(d\psi_\e^{(\le h)})\int P(da^{(\le
h})e^{V^h(\sqrt{Z_h}\psi^{(\le h)}, a^{(\le h)}+A,\l)}, \eea
\end{widetext}
where $P(d\psi_\e^{(h)}) $ has the propagator \be
g^{(h)}_\e(\kk'+{\kk}^\e_F)= {1\over Z_h}\left(\begin{array}{cc}
ik_0 & v_h \O^*(\vec k)
\\ v_h \O(\vec k) & ik_0 \end{array}\right)^{-1}\;\label{v1o}.\ee
$Z_h$ is the wave function renormalization and $v_h$ is the
effective Fermi velocity. Note also that \pref{v1o} can be written
as \be g^{(h)}_\e\sim {1\over Z_h}\left(\begin{array}{cc} ik_0 &
v_h (-i k'_1+\e k_2)
\\ v_h (i k'_1+\e k_2) & ik_0 \end{array}\right)^{-1}\;\label{v1o}
\ee
so that it becomes ever closer to the Dirac propagator. By power
counting, the scaling dimension of the interactions is $D=3-n$ if
$n$ is the number of fields; the local terms quadratic in $a$,
describing the photon mass, are absent by exploiting the Ward
identity generated by \pref{wi}  (see App. E1 of \cite{GMPgauge}
for an explicit computation at one loop ) while the marginal terms
$a^+\partial a$ are vanishing by symmetry; in the same way the
local terms $\psi^+\psi$ are vanishing by parity, and
$\psi^+\partial\psi$ contribute to the wave function
renormalization and the Fermi velocity. The effective potential is
given by (at $\l=0$ for definiteness):
\be
V^h(\sqrt{Z_h}\psi^{\le h},a+A,0)=\sum_\m \sum_\xx Z_{\m,h} (a^{\le h}_{\m,\xx}+A_{\m,\xx}) j^{\le h}_{\m,\xx}+F_h\label{eff},
\ee
 where $F_h$ are the {\it irrelevant terms}, with negative dimension $D<0$.

From the Ward identities \pref{wi} we get
\be
{Z_{0,h}\over Z_h}=1+O(e^2),\quad\quad
{Z_{i,h}\over Z_h v_h}=1+O(e^2)\label{fon}.
\ee
Moreover, by symmetry $Z_{1,h}=Z_{2,h}$ and by an explicit computation,
\be Z_h\sim 2^{-\h h}\quad\quad, v_h\sim 1+A (1-v)2^{\tilde\h h}, \ee
with $\h, \tilde\h>0$ and $O(e^2)$ and $A$ is a bounded function. Therefore the wave function renormalization is diverging (i. e. there is an anomalous dimension),  and  by iterating the RG the Fermi velocity converges to the velocity of light.

The model considered in \cite{GGV} therefore emerges naturally starting from the lattice model. Moreover, the
Lorentz symmetry is restored in the infrared limit, as the Fermi velocity flows up to the velocity of light
(which in our units is $1$). The effective couplings
$e_{h,0}=e {Z_{0,h}\over Z_h}$
$e_{h,i}=e {Z_{1,h} \over Z_h v_h}$ flow to a {\it line} of fixed points
\be
e_{h,i}\to e+O(e^3),
\ee
that is, the theory is {\it exactly marginal}. One obtains then
a renormalized expansion for the current-current correlations
in terms of the effective couplings $e_{\m,h}$, namely
$\s=\s^{(0)}+\s^{(2)}+...$. This differs from perturbative expansion in that  there are no infrared divergences and all coefficients are bounded. We get

\begin{widetext}
\bea\s^{(0)}= N e^2\lim_{\o\to
0^+}\sum_{\e=\pm}\frac1{\o}\sum_{h\le 1}\int\frac{d k_0}{2\p}
\int_{\BBB}\frac{d\vec k'}{(2\pi)^2}\frac{(Z_{1,h})^2}{Z_h Z_{h}}{\rm Tr}\Big\{\s_1 g^{(h)}_\e(\kk')\s_1\big[
g^{(h)}_\e(\kk'+(\o,\vec 0))-
g^{(h)}_\e(\kk')\big]\}\label{fro1},\eea
\end{widetext}
where $\s_j$ are Pauli matrices and $N=2$ in graphene. Note the
presence of a factor ${1\over Z_h}$ for any fermionic line, and of
a factor $Z_{i,h}$ for any vertex. The presence of such factors
could radically alter the conductivity properties; for instance, if
we do not take into account the vertex renormalization, that is we
replace $Z_{i,h}$ with unity, one would get $\s^{(0)}=0$. On the
contrary, thanks to the Ward identity and Eq. \pref{fon} we can
replace $\frac{(Z^{(i)}_{h})^2}{Z_h Z_{h}}$ with $v_h$ up to
$O(e^2)$ terms. The integral then still appears to be non
universal, as it depends of the effective Fermi velocity and on
the lattice details. However it is not so; we can write integral
over the Brillouin zone $\BBB$ as sum of two terms, one $|\O(\vec
k)|\ge \e$ and the other $|\O(\vec k)|\le \e$; the former is
uniformly convergent as $\o\to 0^+$: therefore, we can exchange
the integral with the limit and check that the integral of the
limit is zero simply because the integrand is odd in $k_0$. In the
latter we use \pref{v1o} neglecting the corrections (as the size
of the integral is arbitrarily small); the dependence from $v_h$
disappears through a change of variables and we finally get
$\s^{(0)}={e^2}\frac{N}{8}+O(N e^4)$. Moreover, as shown in
\cite{GMPgauge1}, the low frequency corrections are $O(e^2 \o^2)$.

Let us consider now $\s^{(2)}$; there are three possible contributions, $\s^{(2)}=\s^{(2)}_a+\s^{(2)}_b+\s^{(2)}_c$
but only one of them, which we call $\s^{(2)}_c$, is proportional to $N^2$.
Therefore, if we find it to be non vanishing we can safely conclude that the dc conductivity is different from the non-interacting value,
at least for a generic $N$. The value of such term is

\begin{widetext}
\bea \s^{(2)}_c=-N^2 e^4 \lim_{\o\to 0^+}{1\over 2} \Big\{
{1\over\o} \sum_{\e=\pm}  \sum_{h(\o)\le h\le 1}\int\frac{d
k_0}{2\p} \int_{\BBB}\frac{d\vec k'}{(2\pi)^2}\frac{(Z_{1,
h})^2}{Z_h Z_{h}} {\rm Tr}\Big\{\s_1(\vec k')
g^{(h)}_\e(\kk')\s_1\big[ g^{(h)}_\e(\kk'+(\o,\vec 0))-
g^{(h)}_\e(\kk')\big] \Big\}^2\label{fro} \eea
\end{widetext}
where $h(\o)$ is such that $2^{h(\o)}\sim |\o|$. In writing the above expression we have used that only the renormalized parts of the to
``bubble" diagrams contribute, as their local part is vanishing.
Again using that
$Z_{i,h}/ Z_h=v_h+O(e^2)$ we get $\s^{(2)}_c=-(e^4/2)(N/8)^2+O(e^4 N)$,
and finally \pref{aa} is found. The $N$-dependence allows us to exclude the possibility
of cancelations, and we can conclude that the optical conductivity is different from its non interacting value, with
a leading correction which is universal. Note the difference with the case or Coulomb interactions:
in such a case at one loop $v_h\sim 1/|h|$,
while $Z_h, Z_{i,h}, e_{i,h}$ are essentially constants so that $\s^{(2)}_c$ vanishes. Similarly, for Hubbard interactions the photon propagator
$1/ 2 |\o|$ should be replaced by a constant, and again $\s^{(2)}_c$ would be vanishing.

\section{Effective QED description}

The previous analysis shows that lattice graphene system flows, by iterating the RG, to a fixed point expressed by the the continuum $QED_{4,3}$.
It is useful then to study the conductivity properties in the  continuum theory (assuming as usual that the value of the conductivity depends only on the fixed point of the RG,\cite{bose}) in order to get more information on the subleading corrections.
The action for  the effective model in $2+1$ dimensions is
\bea
&&S = \int d\xx  [  \bar{\Psi}_i (\xx) \gamma_\mu (\partial_\mu - e( a_\mu (\xx) - A_\mu (\xx) ) ) \Psi_i (\xx)\nn\\
&&+ \frac{1}{2}\int d\xx d\yy W_{\mu \nu} ^{-1} (\xx-\yy) a_\mu (\xx) a_\nu (\yy)  ],
\eea
where $\Psi$ is a four-component fermionic field, $i=1,...N$,
$\xx=(x_0, x_1, x_2)$, and the summation convention is assumed. Note that $\vec x=(x_1,x_2)$ is here a continuum variable while
is the previous section was a site on the honeycomb lattice. The bare gauge field propagator is
\begin{equation}
W_{\mu \nu} (\xx) = \frac{1}{2} \int \frac{d^3 \qq}{(2\pi)^3} \frac{e^{i\qq x}}{|\qq|} (\Pi_{\mu \nu} (\qq) + \beta \frac{\qq_\nu \qq_\mu} {\qq^2} ),
\end{equation}
with the usual transverse projector
\begin{equation}
\Pi_{\mu \nu} (\qq) = \delta_{\mu \nu} - \frac{\qq_\mu \qq_\nu}{\qq^2}.
\end{equation}

The above field theory is closely related to the three dimensional quantum electrodynamics ($QED_3$),\cite{appelquist}  with one important caveat: the ``Maxwell" term is now already at the bare level non-analytic in momentum, and proportional to $|\qq|$, and not to the usual $\qq^2$. It can be obtained from the reduced quantum electrodynamics $QED_{4,3}$ in which the electromagnetic fields live in $3+1$ dimensions, but are coupled to fermions which are confined to the lower, $2+1$-dimensional ``brane", by ``integrating out"  the out-of-plane components of the vector potential.\cite{gorbar}  This procedure in general also changes the effective gauge-fixing parameter from $\beta'$ in the original $3+1$-dimensional theory into $\beta$ in the above expression, as in $\beta= (1+\beta')/2$.
 We see that only the Feynman gauge ($\beta=1$) remains invariant under this dimensional reduction, which makes it the most convenient one from the practical point of view. The non-analyticity of the gauge field propagator around $\qq =0$ can be understood as the reason for the exact marginality of the charge coupling. \cite{z}

 Fermions appear quadratically in the action and can be (formally) integrated out. If we redefine the fields as $e a_\mu\rightarrow a_\mu $, and
 $e A_\mu\rightarrow A_\mu $, and then shift the fluctuating field as $a_\mu - A_\mu \rightarrow a_\mu $, the result of this integration would be the
 action
 \begin{widetext}
\bea
\tilde S= \frac{1}{2} \int \frac{d\qq}{(2\pi)^3}\{ ( \frac{N}{8}|\qq| \Pi_{\mu \nu} (\qq) a_\mu(\qq)a_\nu(-\qq)+
\frac{2}{e^2} |\qq|
( \Pi_{\mu \nu} (\qq) + \frac{\qq_\nu \qq_\mu} {\beta \qq^2} ) ( a_\mu(\qq)+ A_\mu (\qq))  ( a_\nu(-\qq)  + A_\nu (-\qq))\}
+V(a)\label{arc}\;.
\eea
\end{widetext}
The first term proportional to $N$ is the familiar one-loop polarization  in the $QED_3$,\cite{qed3} and the second term is the quasi-Maxwell term, which now after the shift of variables also includes the external probe; $V(a)$ is a sum of monomials in the fluctuating gauge field $a$ with degree $\ge 4$.
Note that $V(a)$  does not contain the external probe $A_\mu$ , which after the shift appears only in the quasi-Maxwell term. This allows one to differentiate with respect to the external probe, and so to obtain the current-current correlation function in terms of the {\it exact} gauge field propagator,
$D^{aa}_{\m\n}=\langle a_\m a_\n  \rangle $:
\begin{widetext}
\begin{equation}
\langle j_\mu (\qq) j_\nu (-\qq) \rangle  =  \frac{2}{e^2} |\qq|   ( \Pi_{\mu \nu} (\qq) + \frac{1}{\beta}\frac{\qq_\nu \qq_\mu} {\qq^2} ) -
\frac{2}{e^2} |\qq|   ( \Pi_{\mu \alpha} (\qq) + \frac{1}{\beta} \frac{\qq_\mu \qq_\alpha} {\qq^2} ) D_{\alpha \beta}^{aa} (\qq) \frac{2}{e^2} |\qq|   ( \Pi_{\beta \nu} (\qq) + \frac{1}{\beta} \frac{\qq_\beta \qq_\nu} {\qq^2} ).
\end{equation}
\end{widetext}
Current conservation, on the other hand, dictates that the exact gauge-field propagator has the form:\cite{HT}
\begin{equation}
 D_{\mu \nu}^{aa} (\qq) = \frac{1}{|\qq|} (R \Pi_{\mu \nu} (\qq) + \beta \frac{e^2}{2} \frac{\qq_\nu \qq_\mu} {\qq^2})  ,
\end{equation}
where $R$ is a function of the number of fermions $N$ and of the coupling $e^2$. Inserting this form into the expression for the current-current correlation function, we find that
\begin{equation}
\langle j_\mu (\qq) j_\nu (-\qq) \rangle = \sigma |\qq|   \Pi_{\mu \nu} (\qq),
\end{equation}
where the optical conductivity $\sigma$, in units of $e^2 / \hbar $, and in the limits $T=0$ and $\omega\rightarrow 0$, is simply
\begin{equation}
\sigma = \frac{2}{e^2} ( 1- \frac{2R}{e^2})\;.\label{28}
\end{equation}
Note that the current-current correlation function is completely independent of the gauge-fixing parameter $\beta$, just as one expects. We can rewrite \pref{arc}
as a perturbed Gaussian action
\begin{widetext}
\begin{eqnarray}
\tilde S= \frac{1}{2} \int \frac{d\qq}{(2\pi)^3}\{(\frac{N}{8}+{2\over e^2})|\qq| (\Pi_{\mu \nu} (\qq)  +\g \frac{\qq_\nu \qq_\mu} {\qq^2} )
a_\mu(\qq)  a_\nu (\qq) +  \\ \nonumber
 \frac{2}{e^2} |\qq|   ( \Pi_{\mu \nu} (\qq) + \frac{1}{\beta} \frac{\qq_\nu \qq_\mu} {q^2} )
((2 a_\mu(\qq) A_\nu (-\qq)+A_\m(\qq)A_\n(-\qq))\}\}+V(a)\label{arc1},
\end{eqnarray}
\end{widetext}
where $\g=(2/ \b e^2)(1/ ( N/8+2/e^2)$. In the gaussian approximation, that is neglecting the higher-order terms given by $V(a)$,  the functional integral
can be explicitly performed and the constant $R$ is given by  the value $R_0=((N/8)+(2/e^2))^{-1} $. From \pref{28} we this way find the Ioffe-Larkin-like \cite{IL} result for the conductivity
\begin{equation}
\sigma^{-1}  =[\frac{2}{e^2} ( 1- R_0 \frac{2}{e^2})]^{-1}=
\frac{8}{N} + \frac{e^2}{2}\label{gg1}, 
\end{equation}
which can be interpreted as the addition of the fermion's and the gauge-field's {\it resistivities} into the total resistivity.
Expanding  to the first power in the weak charge coupling $e^2$ yields $\s=(N/8) (1-(N e^2/ 16)+..)$, in agreement with the lattice computation.

To go beyond the gaussian approximation we can expand in powers of the effective charge in the theory $e^2/ ( 2+ (Ne^2 / 8)  ) $; one finds
\be
\frac{1}{R} ={2\over e^2}+{N\over 8}+N x {e^2\over {N\over 8}e^2+2}..., \label{aaa}
\ee
where $x=(92-9\pi^2)/((4\pi)^2 18)$ \cite{teber}. In the weak coupling regime $e^2\ll 1/ N$, after expanding in powers of $e^2$,
\bea
&&\s={2\over e^2}(1-{1\over 1+{N\over 16}e^2+{N x e^4\over 4}+...}) =\nn\\
&&
{N\over 8}(1-{N e^2\over 16}+e^2 4 x+O(e^4)...),
\eea
so that Eq. (1) is recovered,  with the numerical value of the constant
\be
C_1=8x=\frac{23}{(3\pi)^2} - \frac{1}{4}= 0.0089319.
\ee
On the other hand in the large-N limit, $N \gg 1/e^2 $, we expand in powers of $1/N$,
\be
\s={2\over e^2}(1-{2\over e^2} {1\over {N\over 8}+{2\over e^2}+  8 x (1-{16\over N e^2}) +...}),
\ee
so that
\be
\s= \frac{2}{e^2} ( 1- \frac{16}{N e^2}+{128\over N^2 e^4}(2+ C_1 e^2)
+O(N^{-3})).
\ee

\section{Conclusion}

  In conclusion, we have presented the computation of the universal zero temperature, low frequency (optical)  conductivity in the ultimate, relativistic, low-energy regime in graphene, when the Fermi velocity has reached the velocity of light. Although this regime lies beyond presently available experimental conditions, such as the temperature and the sample sizes, the issue of ultimate value of the optical conductivity is theoretically interesting, and presents an important question of principle. We find that the ultimate value of the conductivity is universal, and dependent only on the fine-structure constant of the media surrounding the graphene sheet, but in principle different from the non-interacting value seen in the experiment. The difference from the non interacting value is of the order of the fine structure constant itself, however, and therefore of the relative size of the order of one percent.

\section{Acknowledgement}

IFH was supported by the NSERC of Canada.


\begin{thebibliography}{999999}

\bibitem{N1} R. R. Nair, P. Blake, A. N. Grigorenko, K. S. Novoselov, T. J. Booth, T. Stauber, N. M. R. Peres, A. K. Geim, Science {\bf 320}, 1308 (2008)

\bibitem{L} A. W. W. Ludwig, M. P. A. Fisher, R. Shankar, and G. Grinstein,  Phys. Rev. B {\bf 50}, 7526 (1994).

\bibitem{GMPcond} A. Giuliani, V. Mastropietro and M. Porta, Phys. Rev. B {\bf 83}, 195401 (2011); Comm. Math. Phys. {\bf 311}, 317 (2012).

\bibitem{H1} V. Juri\v ci\' c, O. Vafek and I. F. Herbut, Phys. Rev. B {\bf 82}, 235402 (2010);  I. F. Herbut, V. Juri\v ci\' c, O. Vafek and M. J. Case,  preprint, arXiv:0809.0725.

\bibitem{H0}  I. F. Herbut, V. Juri\v ci\' c and O. Vafek,  Phys. Rev. Lett {\bf 100}, 046403 (2008).

\bibitem{V3} J. Gonz\'alez, F. Guinea and M. A. H. Vozmediano, Phys. Rev. B {\bf 59}, R2474 (1999).

\bibitem{M2} E. G. Mishchenko, Europhys. Lett. {\bf 83}, 17005 (2008).

\bibitem{SS2} D. E. Sheehy and J. Schmalian, Phys. Rev. B {\bf 80}, 193411 (2009).

\bibitem{SF} I. Sodemann and M. M. Fogler, Phys. Rev. B {\bf 86}, 115408 (2012).

\bibitem{RL} B. Rosenstein, M. Lewkowicz, T. Maniv, Phys. Rev. Lett. {\bf 110}, 066602 (2013).

\bibitem{GGV} J. Gonzalez, F. Guinea and M. A. H. Vozmediano, Nucl. Phys. B {\bf 424}, 595 (1994).

\bibitem{GMPgauge} A. Giuliani, V. Mastropietro and M. Porta, Phys. Rev. B {\bf 82}, 121418 (2010); Ann. of Phys. {\bf 327}, 461 (2012).

\bibitem{kane} C. L. Kane and E. J. Mele, Phys. Rev. Lett. {\bf  95}, 226801  (2006).

\bibitem{GMPgauge1} A. Giuliani, V. Mastropietro, Phys. Rev. B {\bf 85}, 045420 (2012).

\bibitem{bose} I. F. Herbut, Phys. Rev. Lett. {\bf 79}, 3502 (1997); ibid. {\bf 81}, 3916 (1998).


\bibitem{appelquist} R. D. Pisarski, Phys. Rev. D {\bf 29}, 2423  (1984);
T. W. Appelquist, D. Nash, L. C. R. Wijewardhana, Phys. Rev. Lett. {\bf 60}, 2575 (1988).

 \bibitem{gorbar} E. V. Gorbar, V. P. Gusynin, and V. P. Miransky, Phys. Rev. D {\bf 64}, 105028 (2001).

 \bibitem{z} I. F. Herbut, Phys. Rev. Lett. {\bf 87},   137004 (2001).

 \bibitem{teber} S. Teber, Phys. Rev. D {\bf 86}, 025005 (2012), particularly Eq. (45a).

 \bibitem{qed3}See, I. F. Herbut, Phys. Rev. Lett. {\bf 88}, 047006 (2002); Phys. Rev. B {\bf 66}, 094504 (2002), and references therein.

 \bibitem{HT} I. F. Herbut and Z. Te\v sanovi\' c, Phys. Rev. Lett. {\bf 76}, 4588 (1996); ibid. {\bf 78}, 980  (1997).

 \bibitem{IL} I. F. Herbut, Phys. Rev. Lett. {\bf 94}, 237001 (2005).

\end{thebibliography}
\end{document}